\documentclass[twoside,twocolumn,9pt]{article}
\usepackage{extsizes}
\usepackage[super,sort&compress,comma]{natbib} 
\usepackage[version=3]{mhchem}
\usepackage[left=1.5cm, right=1.5cm, top=1.785cm, bottom=2.0cm]{geometry}
\usepackage{balance}
\usepackage{times,mathptmx}
\usepackage{sectsty}
\usepackage{graphicx} 
\usepackage{lastpage}
\usepackage[format=plain,justification=justified,singlelinecheck=false,font={stretch=1.125,small,sf},labelfont=bf,labelsep=space]{caption}
\usepackage{float}
\usepackage{fancyhdr}
\usepackage{fnpos}
\usepackage[english]{babel}
\addto{\captionsenglish}{%
	
}
\usepackage{array}
\usepackage{droidsans}
\usepackage{charter}
\usepackage[T1]{fontenc}
\usepackage[usenames,dvipsnames]{xcolor}
\usepackage{setspace}
\usepackage[compact]{titlesec}
\usepackage{hyperref}

\usepackage{multirow}
\usepackage[utf8]{inputenc}
\DeclareUnicodeCharacter{2212}{-}
\DeclareUnicodeCharacter{0327}{}

\usepackage{epstopdf}
\usepackage{color,soul}
\definecolor{cream}{RGB}{222,217,201}

\DeclareUnicodeCharacter{03B2}{+}

\newcommand{\blue}{\textcolor{blue}}

\begin{document}

\pagestyle{fancy}
\thispagestyle{plain}
\fancypagestyle{plain}{
	
}

\makeFNbottom
\makeatletter
\renewcommand\LARGE{\@setfontsize\LARGE{15pt}{17}}
\renewcommand\Large{\@setfontsize\Large{12pt}{14}}
\renewcommand\large{\@setfontsize\large{10pt}{12}}
\renewcommand\footnotesize{\@setfontsize\footnotesize{7pt}{10}}
\makeatother

\renewcommand{\thefootnote}{\fnsymbol{footnote}}
\renewcommand\footnoterule{\vspace*{1pt}%
	\color{cream}\hrule width 3.5in height 0.4pt \color{black}\vspace*{5pt}} 
\setcounter{secnumdepth}{5}

\makeatletter 
\renewcommand\@biblabel[1]{#1}            
\renewcommand\@makefntext[1]%
{\noindent\makebox[0pt][r]{\@thefnmark\,}#1}
\makeatother 
\renewcommand{\figurename}{\small{Fig.}~}
\sectionfont{\sffamily\Large}
\subsectionfont{\normalsize}
\subsubsectionfont{\bf}
\setstretch{1.125} 
\setlength{\skip\footins}{0.8cm}
\setlength{\footnotesep}{0.25cm}
\setlength{\jot}{10pt}
\titlespacing*{\section}{0pt}{4pt}{4pt}
\titlespacing*{\subsection}{0pt}{15pt}{1pt}

\fancyfoot{}
\fancyfoot[LO,RE]{\vspace{-7.1pt}\includegraphics[height=9pt]{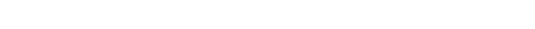}}
\fancyfoot[CO]{\vspace{-7.1pt}\hspace{13.2cm}\includegraphics{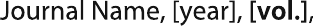}}
\fancyfoot[CE]{\vspace{-7.2pt}\hspace{-14.2cm}\includegraphics{head_foot/RF}}
\fancyfoot[RO]{\footnotesize{\sffamily{1--\pageref{LastPage} ~\textbar  \hspace{2pt}\thepage}}}
\fancyfoot[LE]{\footnotesize{\sffamily{\thepage~\textbar\hspace{3.45cm} 1--\pageref{LastPage}}}}
\fancyhead{}
\renewcommand{\headrulewidth}{0pt} 
\renewcommand{\footrulewidth}{0pt}
\setlength{\arrayrulewidth}{1pt}
\setlength{\columnsep}{6.5mm}
\setlength\bibsep{1pt}

\makeatletter 
\newlength{\figrulesep} 
\setlength{\figrulesep}{0.5\textfloatsep} 

\newcommand{\topfigrule}{\vspace*{-1pt}%
	\noindent{\color{cream}\rule[-\figrulesep]{\columnwidth}{1.5pt}} }

\newcommand{\botfigrule}{\vspace*{-2pt}%
	\noindent{\color{cream}\rule[\figrulesep]{\columnwidth}{1.5pt}} }

\newcommand{\dblfigrule}{\vspace*{-1pt}%
	\noindent{\color{cream}\rule[-\figrulesep]{\textwidth}{1.5pt}} }

\makeatother

\twocolumn[
  \begin{@twocolumnfalse}
\vspace{3cm}
\sffamily
\begin{tabular}{m{4.5cm} p{13.5cm} }

\includegraphics{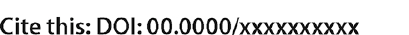} & \noindent\LARGE{\textbf{Effect of Fullerene on domain size and relaxation in a perpendicularly magnetized Pt/Co/C$_{60}$/Pt system}} \\
\vspace{0.3cm} & \vspace{0.3cm} \\

 & \noindent\large{Purbasha Sharangi,\textit{$^{a}$} Aritra Mukhopadhyaya,\textit{$^{b}$} Srijani Mallik\textit{$^{a}$}, Md. Ehesan Ali$^{\dag}$\textit{$^{b}$} and Subhankar Bedanta$^{\ast}$\textit{$^{a}$}} \\

\includegraphics{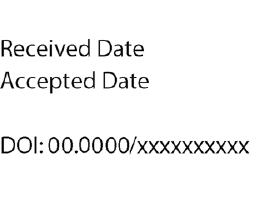} & \noindent\normalsize{Buckminsterfullerene (C$_{60}$) can exhibit ferromagnetism at the interface (called as a spinterface) when it is placed next to a ferromagnet (FM). Formation of such spinterface happens due to orbital hybridization and spin polarized charge transfer at the interface. The spinterface can influence the domain size and dynamics of the organic/ferromagnetic heterostructure. Here, we have performed magnetic domain imaging and studied the relaxation dynamics in Pt/Co/C$_{60}$/Pt system with perpendicular anisotropy. We have compared the results with its parent Pt/Co/Pt system. It is observed that presence of C$_{60}$ in the Pt/Co/Pt system increases the anisotropy and a decrease in the bubble domain size. Further the switching time of Pt/Co/C$_{60}$/Pt system is almost two times faster than Pt/Co/Pt system. We have also performed the spin polarized density functional theory (DFT) calculations to understand the underneath mechanism. DFT results show formation of a spin polarized spinterface which leads to an enhancement in anisotropy.} \\

\end{tabular}

 \end{@twocolumnfalse} \vspace{0.6cm}

  ]

\renewcommand*\rmdefault{bch}\normalfont\upshape
\rmfamily
\section*{}
\vspace{-1cm}


\footnotetext{\textit{$^{a}$~Laboratory for Nanomagnetism and Magnetic Materials (LNMM), School of Physical Sciences, National Institute of Science Education and Research (NISER), HBNI, P.O.- Bhimpur Padanpur, Via-Jatni, 752050, India. E-mail: sbedanta@niser.ac.in}}
\footnotetext{\textit{$^{b}$~Institute of Nano Science and Technology, Knowledge City, Sector-81, Mohali, Punjab 140306, India. E-mail: ehesan.ali@inst.ac.in }}

\footnotetext{\dag~Electronic Supplementary Information (ESI) available.}


\section{Introduction}

Organic spintronics is an emerging research topic in the last two decades due to exciting physical phenomena as well as its potential in various spintronic applications. Organic semiconductors (OSCs) have drawn immense research interest for spintronic applications due to low spin orbit coupling, less hyperfine interaction, long spin lifetime, low cost and mechanical flexibility \cite{xiong2004giant,dediu2009spin,atodiresei2010design,barraud2010unravelling}. Spin valve like structure using OSC as a spacer layer has already been shown to exhibit high magnetoresistance \cite{ding2017inverse,zhang2015magnetoresistance,li2013effect,dediu2008room}. It has been reported that the performance of the spin valve device depends on the interface of the OSC and ferromagnet (FM) \cite{atodiresei2010design,barraud2010unravelling,wang2016effect,tran2013magnetic,friedrich2015chemically,zhang2011interface,callsen2013magnetic,wang2013peculiarities,lach2012metal}. There is a high chance of formation of a spinterface due to spin polarized charge transfer and orbital hybridization at the OSC-FM interface \cite{sanvito2010rise,djeghloul2013direct}. Because of this the density of states of the OSC get modified and the OSC can exhibit ferromagnetism \cite{tran2013magnetic,moorsom2014spin,tran2011hybridization,mallik2018effect,djeghloul2016high,mallik2019tuning,mallik2019enhanced,arruda2020surface,arruda2019modifying}. The effect of spinterface has been shown in a few in-plane magnetized systems \cite{moorsom2014spin,tran2011hybridization,mallik2018effect,djeghloul2016high,mallik2019tuning,mallik2019enhanced}. Recently we have shown that due to the spinterface the magnetization reversal and domains are modified in an epitaxial MgO (100)/Fe/C$_{60}$ system \cite{mallik2018effect}. The magnetic moment per C$_{60}$ cage was found to be $\sim$ 2.95 $\mu_B$ \cite{mallik2018effect}. Similarly, for C$_{60}$ deposited on a polycrystalline Fe films also exhibited spinterface and domain size was reduced \cite{mallik2019tuning}. Further in another study we have shown that the spinterface in a Co/C$_{60}$ system has similar effect, however the anisotropy increased due to the spinterface \cite{mallik2019enhanced}. Although there have been a few works on in-plane magnetized films however the study of such spinterface on domains in a perpendicularly magnetized films are scarce.

 \begin{figure*}
	\centering
	\includegraphics[width=1.0\linewidth]{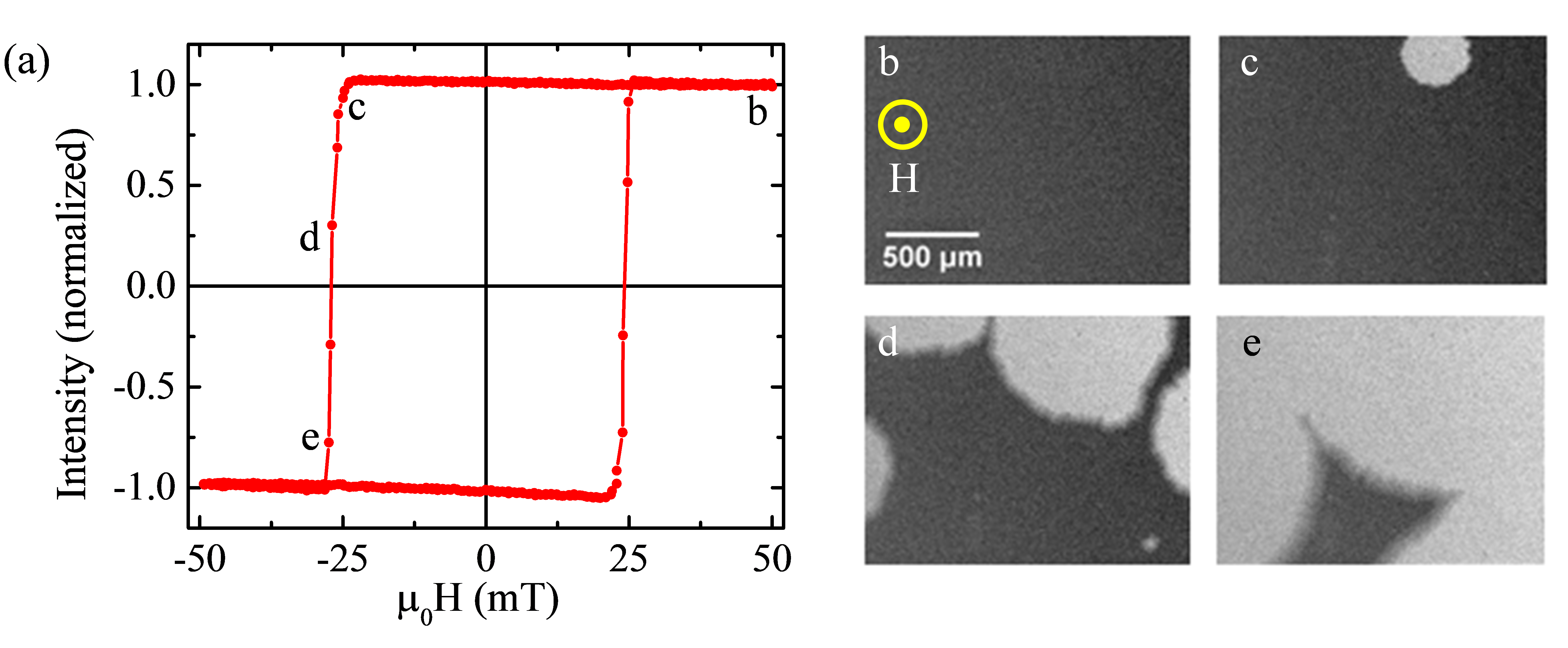}
	\caption{(a) hysteresis loop measured by MOKE based microscopy in polar mode for sample 1. (b - e) are the corresponding domain images at magnetic fields of 50 ,-24.73,-26.89 and-27.85 mT, respectively.}
	\label{fig:fig-1}
\end{figure*}

Perpendicular magnetic anisotropic (PMA) systems are the most suitable candidates for data storage devices due to their high intrinsic anisotropy \cite{nishimura2002magnetic}. Due to high thermal stability and low energy consumption the PMA based devices have advantages over the in-plane ones. In systems with PMA, magnetic moments are aligned perpendicular to the film plane which enhances the spin-flipping efficiency thereby reducing the required current density which may be useful for spin-orbit torque (SOT) based devices \cite{nishimura2002magnetic,mangin2006current}. Bairagi et al. demonstrated that with increasing the thickness of the OSC (C$_{60}$) layer on a Co ultrathin film one can tune the anisotropy of the system from in-plane to out-of-plane \cite{bairagi2015tuning,bairagi2018experimental}. It has been also shown that a C$_{60}$ layer can enhance the PMA of the Ni thin film in a Ni/C$_{60}$ bilayer system \cite{pang2016manipulating}. Further it has been observed that adsorbing the organic molecules on CoFe$_{3}$N surface enhance the PMA of the system \cite{li2018orbital}. Therefore, the study of domain and magnetic relaxation is quite appealing in FM-OSC systems with PMA. From application point of view domain engineering via various approaches are appealing. For example, recently we have shown that by making magnetic antidot lattices (MALs) in a Pt/Co/Pt film, the domain size was significantly reduced \cite{mallick2018relaxation}. However, domain engineering via spinterface and in particular in Pt/Co/Pt system has not been studied so far. 

In this paper, we have considered a Pt/Co/Pt system which exhibits high PMA. We have studied a Pt/Co/C$_{60}$/Pt system in which the effect of spinterface at the Co-C$_{60}$ interface on magnetic domain size and relaxation have been investigated. The formation of spinterface and its effect on the local magnetic environment has been studied by means of density functional theory (DFT) calculations. We have observed that due to the spinterface the size of bubble domains gets significantly reduced. Further from the magnetization relaxation measurements it is found that the relaxation time decreases for the Pt/Co/C$_{60}$/Pt system.

\begin{figure*}
	\centering
	\includegraphics[width=1.0\linewidth]{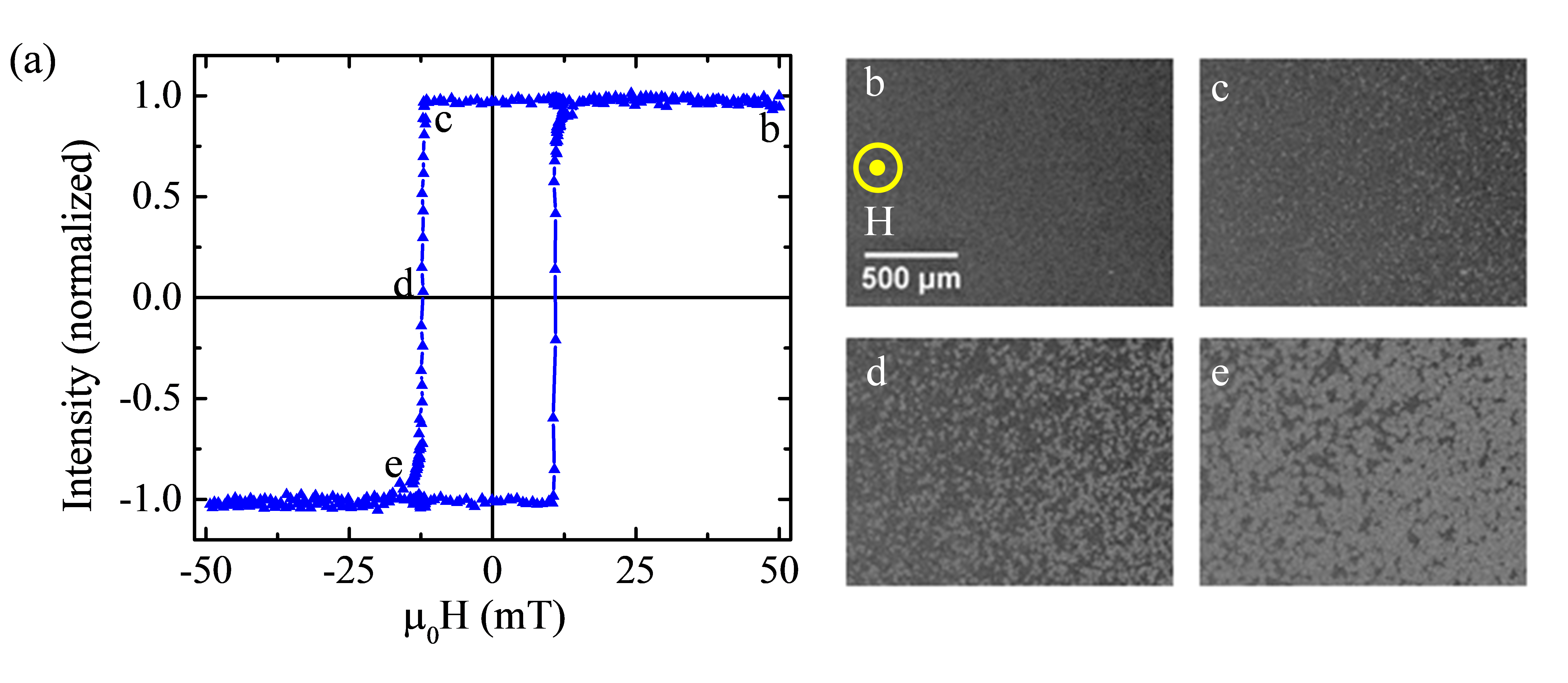}
	\caption{(a) hysteresis loop measured by MOKE based microscopy in polar mode for sample 2. (b - e) are the corresponding domain images at magnetic fields of -50,-11.72,-12.10 and-12.30 mT, respectively.}
	\label{fig:fig-2}
\end{figure*}

\section{Experimental details:}
We have prepared two Pt/Co/Pt thin films with and without C$_{60}$. The sample structures are the following:
\\Sample 1: Si/ Ta (5.5 nm)/Pt (4.1 nm)/Co (0.8 nm)/Pt (4.6 nm) and
\\Sample 2: Si/ Ta (5.5 nm)/Pt (4.1 nm)/Co (0.8 nm)/C$_{60}$ (1.7 nm)/Pt (4.6 nm)
\\The thin films have been prepared using DC sputtering (Co, Ta), RF sputtering (Pt) and thermal evaporation (C$_{60}$) techniques in a multi-deposition high vacuum chamber manufactured by Mantis Deposition Ltd., UK. The base pressure in the chamber was $5\times 10^{-8}$ mbar. All the layers were deposited in-situ to avoid surface contamination and oxidation. Ta was used as a seed layer to promote the (111) direction growth of Pt. Further to maintain the growth uniformity, we have rotated the substrate at 20 rpm during deposition of all the layers. To prevent from oxidation of the films a Pt capping layer of 4.6 nm thickness was deposited.
We have measured the magnetic domains and relaxations at room temperature by magneto-optic Kerr effect (MOKE) based microscopy manufactured by Evico magnetics GmbH, Germany. The measurements were performed in polar mode where the magnetization is parallel to the plane of incidence and perpendicular to the sample surface.

The density functional theory (DFT) calculations has been performed using Vienna Ab-initio Simulation Package (VASP) \cite{kresse1996efficient} in order to understand the atomistic details and the electronic structure of the spinterface. In the calculations the valence electronic states are expanded with a plane wave basis set, while the core electrons are treated with pseudopotential. The valence-core interaction is represented by full-potential Projected Augmented Wave (PAW) method \cite{kresse1999ultrasoft}. The Generalized Gradient Approximations (GGA) is used to treat the exchange-correlation potentials with the Perdew, Bruke, and Ernzerof (PBE) functional \cite{perdew1996generalized}. A plane-wave cut-off of 500 eV is used to guarantee a good convergence of the total energy. The convergence tolerance for the self-consistent electronic minimization is set to $10^{-6}$ eV/cycle. For optimization a Gaussian smearing parameter of 0.2 eV is used to smear the bands. A Monkhorst-Pack K-points grid is taken as (3$\times$3$\times$1) for the surface slabs. A vacuum of 10 \AA\ is added to the z-direction of the surface super cells to ensure no interaction between periodic images at that particular direction. The Co 3d orbitals are treated with a U$_{eff}$=3.0 eV for the calculation of magnetic anisotropy energy (MAE) and exchange coupling. Spin-orbit coupling has also been included in our calculations for evaluating the sub-meV MAEs. The surface slabs is constructed using optimized cell parameters $a$=2.50 \AA\ and $c/a$=1.6 for hexagonal closed packed (HCP) Co and $a$=2.48 \AA\ for face centered cubic (FCC) Co.

\section{Results and discussion:}

\begin{figure*}[hbt!]
	\centering
	\includegraphics[width=1.0\linewidth]{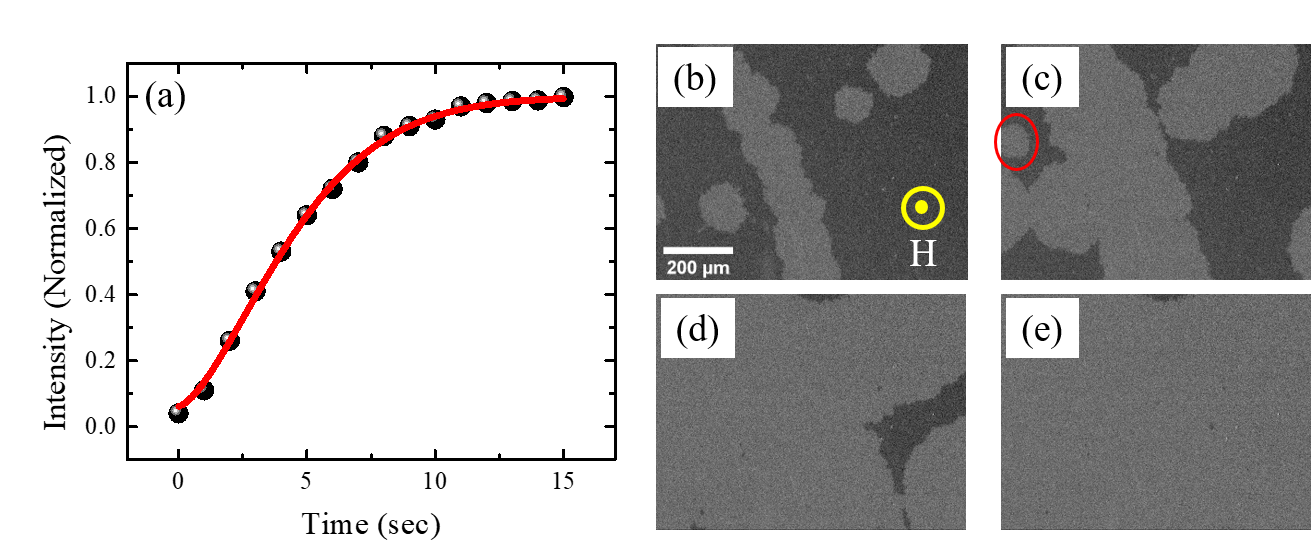}
	\caption{(a) Relaxation data for sample 1 measured at $H_{M}$=0.97 $H_{C}$, where the solid circles represent the raw data and red line is the best fit using compressed exponential function given as equation (1). (b) - (e) represent the domain images captured at 0, 3, 8, 15 sec respectively. The scale bar shown in (b) is same for all the images and the applied field is out of the plane. New domain nucleation is marked in red circle (c).  }
	\label{fig:fig-3}
\end{figure*}
\begin{figure*}
	\centering
	\includegraphics[width=1.0\linewidth]{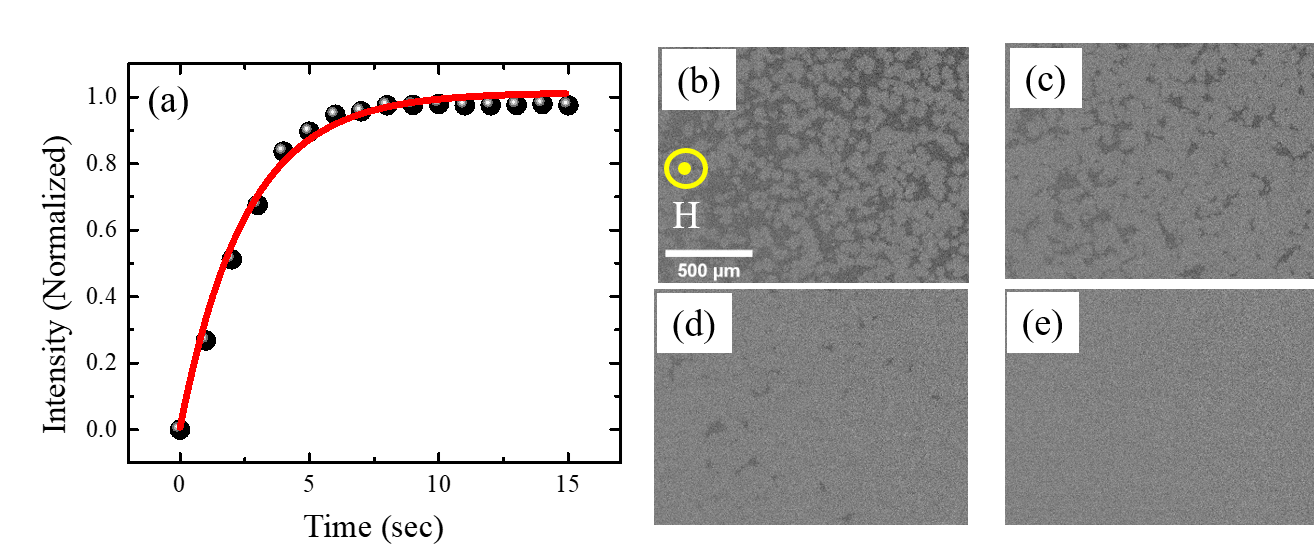}
	\caption{(a) Relaxation data for sample 2 measured at  $H_{M}$=0.97 $H_{C}$, where the solid circles are representing the raw data and red line is the best fit using compressed exponential function given in equation (1). (b) - (e) represent the domain images captured at 0, 2, 4, 15 sec respectively. The scale bar shown in (b) is same for all the images and the applied field is out of the plane. }
	\label{fig:fig-4}
\end{figure*}

The thickness of the Co layer is very crucial for PMA system. In order to know the exact thickness and roughness of all the layers, we have performed X-ray reflectivity (XRR) measurement and fitted the data using GenX software. The XRR data and their best fits are shown in supplementary information figure S1. It has been observed that the thickness of the Co layer is same for both the Pt/Co/Pt and Pt/Co/C$_{60}$/Pt samples (refer to the Table S1 in supplementary information). 
\\Figure 1 (a) shows the hysteresis loop measured by MOKE based microscopy in polar mode for sample 1. The coercive field ($H_{C}$) is 25.56 mT. Figure 1 (b-e) show the domain images of sample 1 at positive saturation field (+$H_{S}$), nucleation field ($H_{N}$), near to coercive field ($H_{C}$) and negative saturation field (-$H_{S}$), respectively. Bubble domains are observed in the samples as the anisotropy ratio, $Q$=$K_{u}$/$K_{d}$ $\gg$ 1, where  $K_{u}$ and  $K_{d}$ are perpendicular anisotropy and stray field energy densities, respectively \cite{hubert2008magnetic}. With increasing the magnetic field bubble domains expand and at saturation field they merge with each other. Figure 2 (a) shows hysteresis loop and figure (b-d) show the domain images of sample 2 at +$H_{S}$, $H_{N}$, $H_{C}$, and -$H_{S}$, respectively. It is clearly observed that the size of the domains for sample 2 become smaller in comparison to sample 1. Also, the coercivity for sample 2 (11.55 mT) is less than that of sample 1. Further, using SQUID magnetometry we have calculated the magnetic anisotropy (shown in supplementary figure S2). It has been found that anisotropy is more for sample S2 (with C$_{60}$ layer) in compariosn to S1(without C$_{60}$). The possible reason behind it may be the formation of spinterface at Co-C$_{60}$ interface which increases the anisotropy as well as the decrease of domain size of the system. However, the overall nature of the bubble domains remains same like its parent Pt/Co/Pt thin film. 

Depending on the domain dynamics, switching speed of spintronic devices can be tuned. To calculate the relaxation time of a system we have performed magnetization relaxation measurement using Kerr microscopy at room temperature. Here first we have saturated the sample and then reversed the magnetic field to a sub-coercive field (0.93 $H_{C}$, 0.95 $H_{C}$ etc.) and kept the field constant. The magnetization will relax with time under a constant Zeeman energy and complete the reversal process via domain nucleation and/or domain wall (DW) motion under the influence of thermal activation energy. Domain images were captured in a regular interval of time during this process.

We have calculated average normalized intensity using ImageJ software and plotted it with respect to time. The relaxation curve can be fitted using various models \cite{fatuzzo1962theoretical,labrune1989time,adjanoh2011compressed,mallick2015effect,mallick2015size,chowdhury2016study}. Here we have fitted our experimental data using compressed exponential function \cite{mallick2018relaxation}: 

\begin{equation}
	I(t)=I_{1} +I_{2}(1-exp(-(\frac{t}{\tau})^\beta))
\end{equation}

where $I(t)$ represents the Kerr intensity at time t, $I_{1} + I_{2}$ is normalized Kerr intensity measured at saturation,  $\tau$ is relaxation time constant and $\beta$ is an exponent having value between 1 (domain nucleation dominated magnetization reversal) and 3 (dominated by DW motion).

\begin{figure*}
	\centering
	\includegraphics[width=1.0\linewidth]{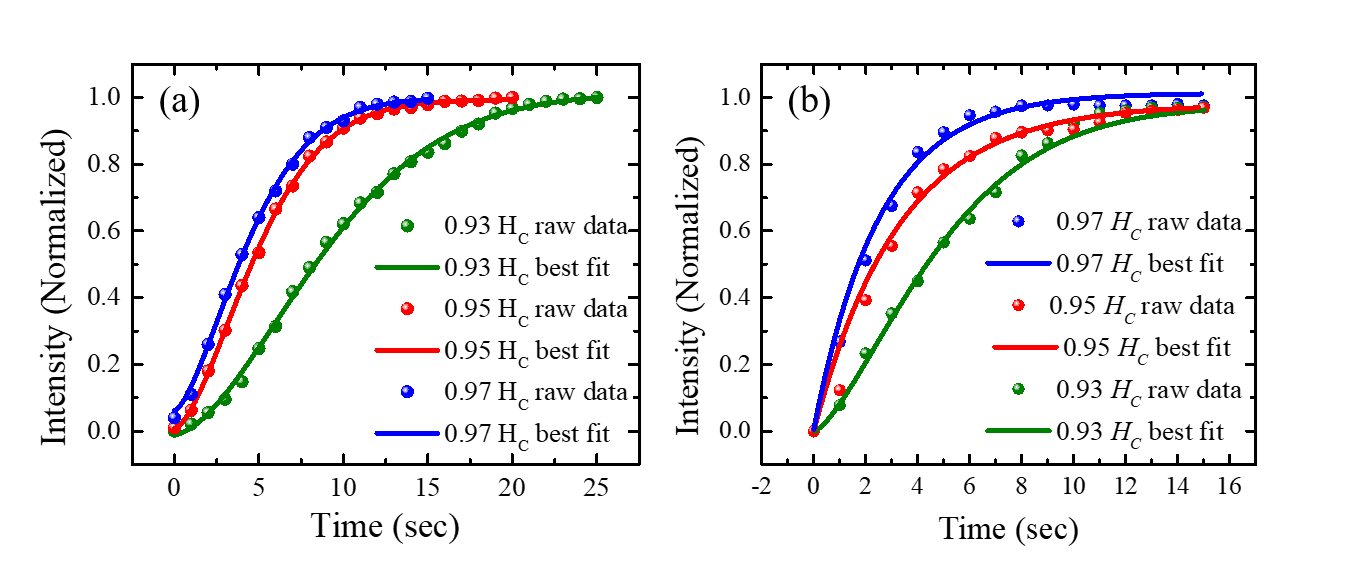}
	\caption{ Relaxation data measured at $H_{M}$= 0.97 $H_{C}$ (blue solid circles), $H_{M}$= 0.95 $H_{C}$ (red solid circles), $H_{M}$= 0.93 $H_{C}$ (green solid circles) for sample 1 (a) and sample 2 (b). The solid lines represent the best fits using the compressed exponential function given in equation (1).} 	\label{fig:fig-5}
\end{figure*}

\begin{table*}
	\centering
	\caption {Parameters obtained from the best fits of the data shown in Fig. 5 using compressed exponential function equation (1)}
	
	\begin{tabular}{|c|c|c|c|c|c|c|}
		\hline
		\multirow{2}{*}{Sample   name} & \multicolumn{2}{c|}{0.97Hc} & \multicolumn{2}{c|}{0.95Hc} & \multicolumn{2}{c|}{0.93Hc} \\ \cline{2-7} 
		& $\tau$          & $\beta$         & $\tau$          & $\beta$         & $\tau$           & $\beta$        \\ \hline
		Sample   1                     & 5.15 ± 0.03  & 1.53 ± 0.05  & 5.75 ± 0.05  & 1.59 ± 0.03  & 10.42 ± 0.14  & 1.66 ± 0.05 \\ \hline
		Sample   2                     & 2.75 ± 0.12  & 1.07 ± 0.05  & 3.74 ± 0.04  & 1.13 ± 0.12  & 5.98 ± 0.19   & 1.37 ± 0.07 \\ \hline 
	\end{tabular}
\end{table*}

\begin{figure*}
	\centering
	\includegraphics[width=0.8\linewidth]{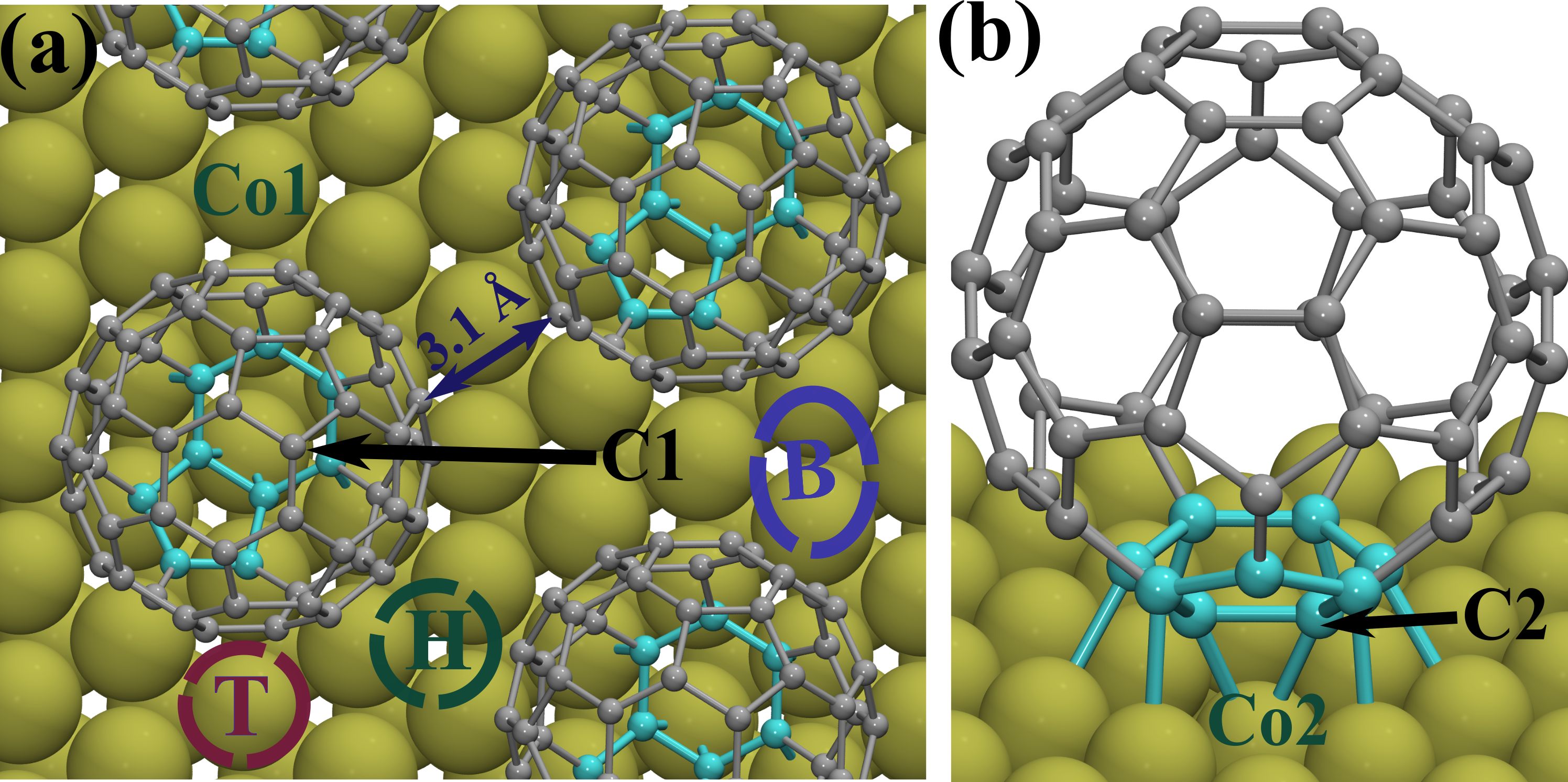}
	\caption{Chemisorption of C$_{60}$ molecule with a pentagonal-hexagonal adsorption geometry on HCP-Co(0001) substrate. (a) shows the top view of extended simulation supercell. (b) shows the spinterface formation on a single C$_{60}$ unit. The relatively bigger tan-coloured balls represent Cobalt atoms whereas the grey balls represent Carbon atoms. Cyan balls highlight the C atoms from the pentagon-hexagon which take part in chemical bond formation with the substrate Co atoms. The circled areas on the surface Co layer represent different adsorption sites on the Co substrate. T, H and B stands for Top, Hollow and Bridged adsorption sites. 4 atoms have been labelled; Co1 and C1 are the free Cobalt and Carbon atoms. Co2 is the surface Cobalt atom which is bonded to the C2 Carbon atom of the $C_{60}$.} 	\label{fig:fig-6}
\end{figure*}

\begin{figure}[hbt!]
	\centering
	\includegraphics[width=8.6 cm]{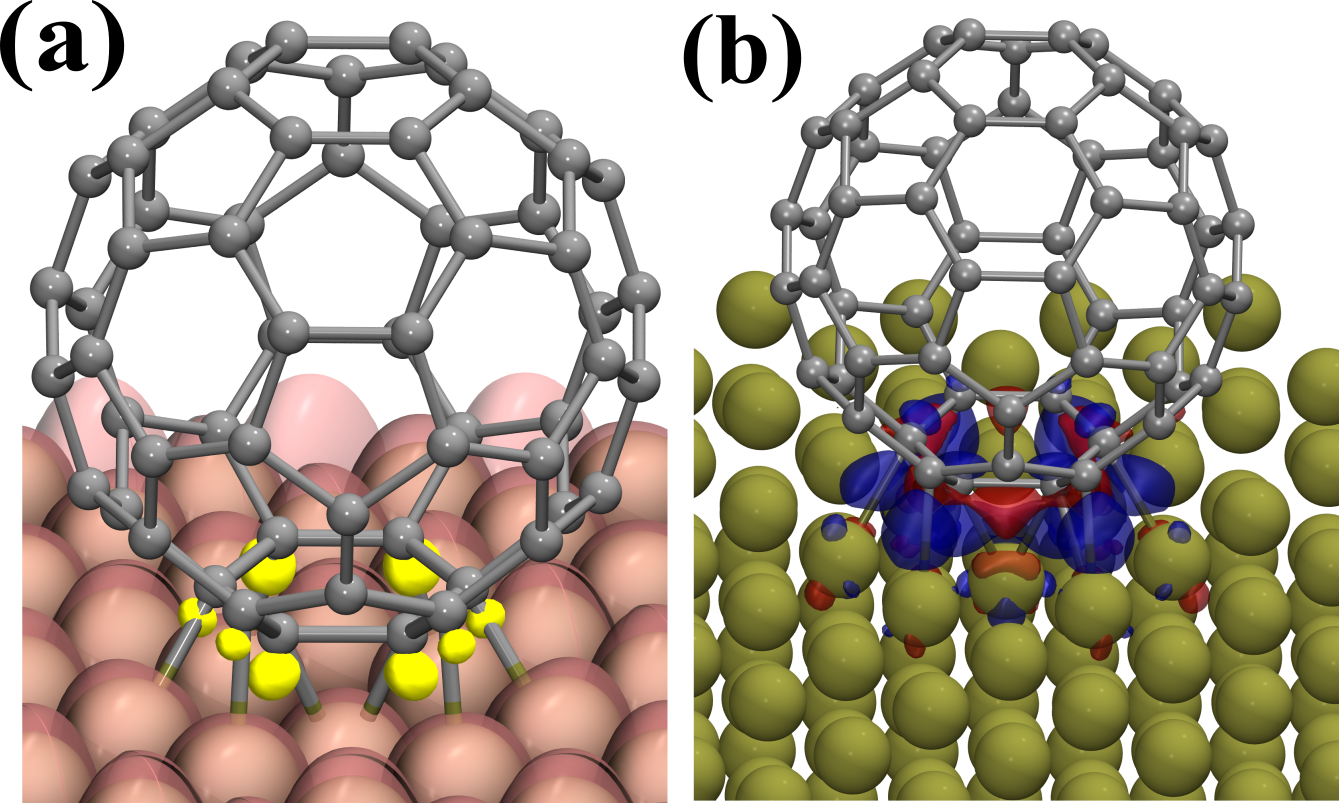}
	\caption{Formation of the spinterface at the HCP-Co(0001) substrate and pentagonal-hexagonal C$_{60}$ interface. (a) Distribution of spin-density at the spinterface in which pink isosurface represents majority spin density while the yellow isosurface represents minority spin density. (b) The charge  density redistribution at the spinterface. The difference density has been obtained as $\rho_{diff}=\rho_{Co+C_{60}}-\rho_{Co}-\rho_{C_{60}}$, where $\rho$ represents the $e^-$-density for the corresponding chemical entity. Red isosurface indicates the depletion of $e^-$ density while blue isosurface indicates the accumulations. The isosurfaces are mapped with an iso-value of 0.04 $e^-$/\AA$^3$}
	\label{fig:fig7}
\end{figure}

\begin{figure*}[hbt!]
	\centering
	\includegraphics[width=17 cm]{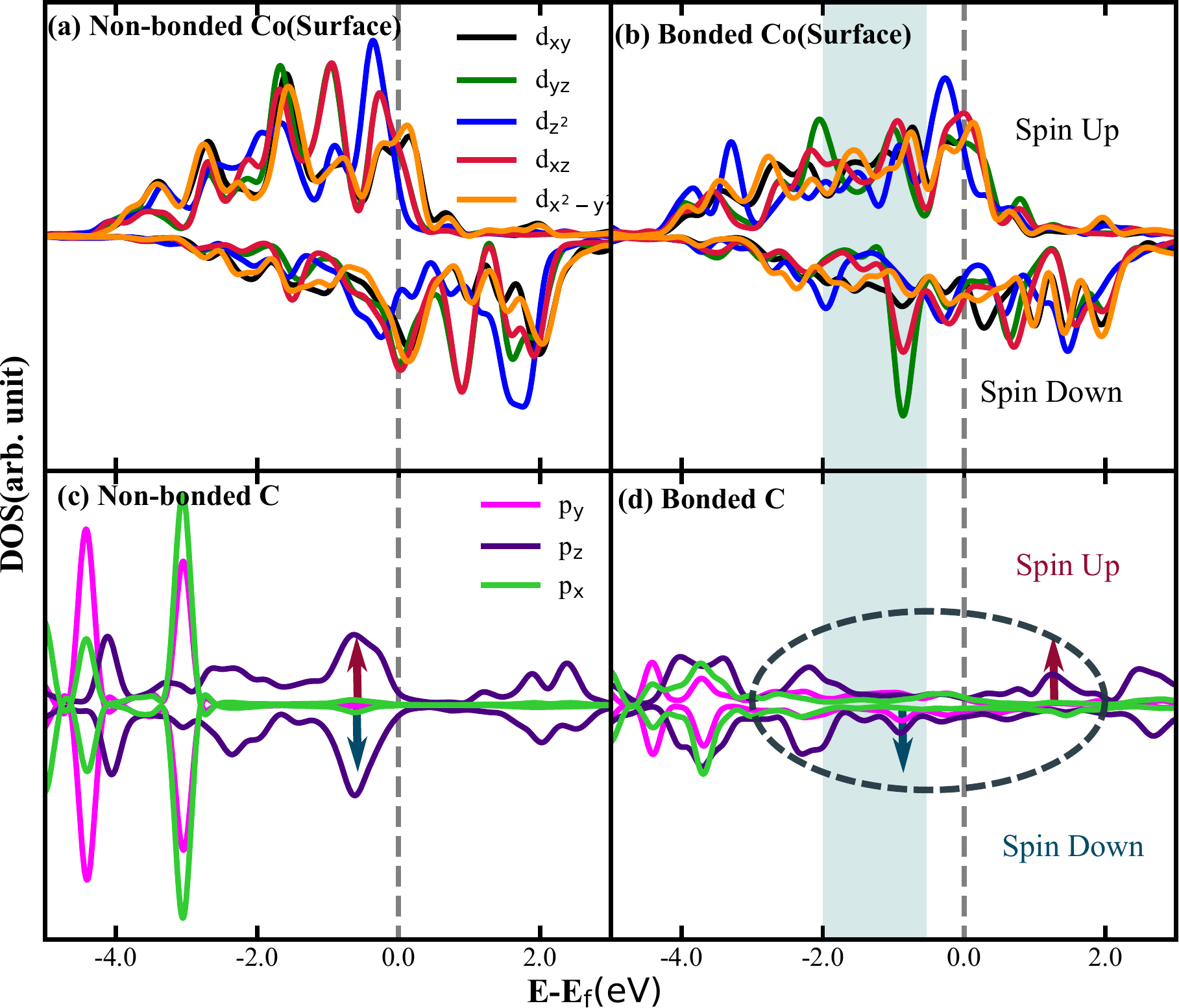}
	\caption{Spin polarized lm-resolved density of states for different Co(3d) and C(2p) atoms at the spinterface of HCP-Co (0001) substrate in Penta-Hexa C$_{60}$ as specified in Fig. {\ref{fig:fig-6}}. (a) Non-bonded Co atom at the surface layer of the Co-substrate. (b) Bonded Co atom at the surface layer. (c) Non-bonded C atom which resides in opposite to the spinterface region. (d) C atom which is bonded to Co atom described in (b). These specific atomic sites are marked in fig. 6. The cyan shaded region depicts the energy at which the bonded Co and C atoms' states overlap to form the hybridised state. The area within the dashed line in (d) represents the spin split states of the C atom p-orbitals which forms as a consequence of hybridization with the $d_{xz}$, $d_{yz}$ and $d_{z^2}$ orbitals of the bonded Co atom. The energies at which spin split states of the C atoms have formed, are marked with the arrows in (c) and (d).}
	\label{fig:fig-8}
\end{figure*}
\begin{table*}
	\begin{center}

		\caption {Parameters used to define the interface formation between different Co-slabs and C$_{60}$}
		\begin{tabular}{|c|c|c|c|c|}
			\hline
			\multicolumn{2}{|c|}{System}                    & $E_{ad}$(kcal/mol)$^{a}$ & $e^-$-transferred$^{b}$ & Number of Co-C bond$^{c}$ \\ \hline
			\multirow{2}{*}{HCP-Co(0001)} & Penta- C$_{60}$      & -94.89         & 1.50             & 9                    \\ \cline{2-5} 
			& Penta-Hexa- C$_{60}$ & -100.30        & 1.41             & 8                    \\ \hline
			\multirow{2}{*}{FCC-Co(111)}  & Penta- C$_{60}$      & -83.15         & 1.45             & 7                    \\ \cline{2-5} 
			& Penta-Hexa C$_{60}$  & -95.10         & 1.39             & 8                    \\ \hline
		\end{tabular}
	\end{center}
	\textsuperscript{a} Adsorption Energy, $E_{ad}$=$E_{Co+C_{60}}-(E_{Co}+E_{C_{60}})$\\
	\textsuperscript{b} e$^{-}$ - transferred from Co substrate to C$_{60}$ = Bader population on C$_{60}$ in Co+C$_{60}$ – Bader population on C$_{60}$.\\
	\textsuperscript{c} Co-C bond distance Cut-off = 2.2 $\AA$
\end{table*}

\begin{table*}
	\centering
	\caption {Computed relative energies of the antiferromagnetic (AFM-i) spin configuration where one magnetic moment is antiferro-magnetically coupled with the rest of the substrate atoms. $\Delta E$  has been calculated as $\Delta E = E_{AFM-i} - E_{FM}$. For surface all the magnetic moments are identical and hence only one relative energy is there. The MAE has been calculated here, as MAE = $E_{axis} - E_{Z-axis}$. Positive MAE value indicates out of plane magnetic anisotropy. }
	\begin{tabular}{|c|c|c|c|c|c|}
		\hline
		\multirow{2}{*}{System} & \multicolumn{2}{c|}{MAE(meV)} & \multicolumn{3}{c|}{$\Delta E$(meV)*}        \\ \cline{2-6} 
		& X-axis        & Y-axis        & AFM-1(1.9) & AFM-2(1.8) & AFM-3(1.6) \\ \hline
		HCP-Co(0001)            & 8.25          & 8.26          & 661.44     &            &            \\ \hline
		+Penta-C$_{60}$              & 7.55          & 7.37          & 935.06     & 698.16     & 626.39     \\ \hline
		+Penta-Hexa-C$_{60}$         & 8.94          & 8.78          & 916.15     & 660.50     & 717.04     \\ \hline
		FCC-Co(111)             & 0.59          & 0.58          & 642.33     &            &            \\ \hline
		+Penta-C$_{60}$              & 0.91          & 0.63          & 928.64     & 673.75     & 694.67     \\ \hline
		+Penta-Hexa-C$_{60}$         & 2.00          & 1.98          & 914.67     & 673.90     & 686.08     \\ \hline
	\end{tabular}
	\\ \textsuperscript{*} Values inside the parenthesis represent the approximate magnetic moments in $\mu_B$ (DFT+U) of the corresponding ferromagnetic configuration which has been inversed in a particular AFM configuration.
\end{table*}

Figure 3 (a) shows the relaxation behaviour of Pt/Co/Pt thin film (sample 1) at $H_{M}$=0.97 $H_{C}$ and (b – e) show the domain images captured at 0, 3, 8, 15 seconds, respectively, which are marked in (a). Figure 3(b) has been captured at $t$ = 0, i.e. just after applying the constant magnetic field ($H_{M}$). After that domains are evolved with time under the influence of thermal activation energy. From the fitting of the relaxation curve using equation (1), we have found $\beta$ = 1.55 $\pm$ 0.05, which ensures that the magnetization reversal started via domain nucleation and subsequently via DW motion. \cite{mallick2018relaxation}.
Similarly, figure 4 (a) shows the relaxation behaviour of Pt/Co/C$_{60}$/Pt thin film (sample 2) at $H_{M}$=0.97 $H_{C}$ and (b – e) show the domain images captured at 0, 2, 4, 15 seconds, respectively, which are marked in (a). By fitting the data to equation (1), the $\beta$ value is found to be 1.07  $\pm$ 0.05 which also indicates domain nucleation dominated magnetization reversal\cite{mallick2018relaxation}. 

To study the speed of the relaxation with different values of the Zeeman energy, we have performed the relaxation measurements at various sub-coercive fields (0.93, 0.95, 0.97 $H_{C}$) for both the samples 1 and 2. Figures 5 (a) and (b) show the relaxation behaviour for samples 1 and 2, respectively, measured at $H_{M}$= 0.97 $H_{C}$ (blue squares), $H_{M}$= 0.95 $H_{C}$ (red triangles), $H_{M}$= 0.93 $H_{C}$ (green circles). The fitting exponent $\beta$ and relaxation time ($\tau$) values obtained from the best fits to equation (1) for both the samples are given in Table 1. For both the samples, $\tau$ decreases for increasing HM, which is expected. However, for Pt/Co/C$_{60}$/Pt thin film (sample 2) the relaxation time is reduced by $\sim$ 50 \% (i.e. the relaxation process is faster) as compared to Pt/Co/Pt thin film (sample 1). Therefore, it is inferred that by introducing a C$_{60}$ layer the switching speed of a system can be tuned which is a promising way to build device applications.

The nature of the spinterface at the Co-C$_{60}$ interface has been explored applying DFT calculations. Two different types of Co-substrates are considered here; viz, HCP-Co (0001) and FCC-Co (111), as the exact morphology of the experimental samples are not known. Two different sites of C$_{60}$ were deposited on the Co-substrates terming them as 'pentagonal' and 'pentagonal-hexagonal' configurations {\cite{moorsom2014spin,bairagi2015tuning,bairagi2018experimental}. We have used a 3 layer Co slab(in conjunction with the experimental samples) with a 4$\times$4 in plane repetition. For the  pentagonal and hexagonal adsorption, the starting geometry for optimization is taken in such a way that the shared C-C bond between pentagon and hexagon is on the top of a surface Co atom and both the pentagon and hexagon faces} the Co substrates. For pentagonal adsorption only C atom from the pentagon faced the Co substrate. The optimized Co (HCP001) and pentagonal-hexagonal C$_{60}$ interface has been shown in figure 6. The formation of such spinterface amicably modifies the isotropic magnetic exchange interactions of the surface atoms as well as amends the magnetic anisotropy of the substrates.  The quantitative estimations of such interfacial magnetic properties are obtained from our  GGA+\textit{U}+SOC calculations.

A quantitative description of different spinterface formation of the above mentioned systems has been provided in Table 2. Comparison of $E_{ads}$ suggest a strong chemisorption for all systems and the penta-hexa adsorption is favoured over pentagonal variant for both type of Co substrates. The spin-polarization at the interface is shown in Fig. 7(a), where the C$_{60}$ is inversely spin polarized as compared to the Co-substrate. Bader populatin analysis{\cite{henkelman2006fast}} (Table 2) suggest a depletion of electronic population ($\sim$ 1.4 $e^-$) from the C$_{60}$ molecule. In Fig. 7(b) the electron density redistribution due to formation of chemisorbed spinterface has been graphically represented. It depicts how the $\pi$ electron density of the adsorbed C atoms get involved in the bond formation with the substrate Co atoms. The change in the electronic structure at the spinterface has been investigated through the site projected $lm$-resolved density of states ( Partial DOS projected on orbital($l$) and magnetic($m$) quantum numbers ) in fig. 8. The out of plane $d_{z^2}$, $d_{xz}$ and $d_{yz}$  orbitals of surface Co and p-orbitals of the bonded C atom changes significantly over a broad energy region near Fermi level, indicating formation of a strongly hybridized state between the bonded atoms. The spin split state shown in figure 8(d) is mostly composed of the $p_z$ atomic orbital of the bonded C atom. This indicates involvement of $\pi$-molecular orbital of the C$_{60}$ with the out of plane $d$-orbitals of the Co substrate atoms in the formation of hybridised spin split states. This implies a $p_{\pi}-d_{\pi}$ interaction between the chemisorbed C$_{60}$ and the Co substrate.

There are mainly three types of magnetic moment present at the interfacial Co layer (Supplementary information figures S3 and S4).  In order to evaluate the change in local exchange interactions at the interfacial layer of the Co-substrate,  we have considered three different antiferromagnetic spin configuration where the spin moment of one of the atoms with a particular magnetic moment is reversed.   The energy difference between the corresponding antiferromagnetic configuration(denoted as AFM-i in the Table 3) and the ground state ferromagnetic configuration  is tabulated in Table 3.  

There are mainly three types of magnetic moment present at the interfacial Co layer (Supplementary information figures S3 and S4). In order to evaluate the change in local exchange interactions at the interfacial layer of the Co-substrate,  we have considered three different antiferromagnetic (denoted as AFM-i in Table 3) spin configuration where the spin moment of one of the Co atoms with a particular magnetic moment is reversed.  The energy difference between the corresponding antiferromagnetic configuration and the ground state ferromagnetic configuration(tabulated as $\Delta E$ in Table 3) indicates how the local magnetic exchange interactions are being modified due to the formation of the spinterface as compared to the clean surface. For most of the configurations the magnetic exchange interactions are enhanced. This reveals that the local magnetic environment at the adsorption site become discrete from rest of the ferromagnetic surface {\cite{callsen2013magnetic,raman2013focusing}}. Evidently these modified magnetic environments regulate the nature of magnetic domains in the sample with C$_{60}$.    

The magnetic anisotropy for the interfaced systems is calculated accounting the spin-orbit couplings in the non-collinear magnetic calculations. The magnetic easy-axis is aligned out of plane i.e, perpendicular to surface layers for all the Co-surfaces studied here. The values of the magnetic anisotropy energy (MAE) w.r.t. to the easy axis anisotropy for pristine Co surface as well as for C$_{60}$ adsorbed surface are compared in Table \blue{3}.  It is evident that there are enhancements in MAE (except for penta-C$_{60}$ on HCP-Co which is less favorable configuration) due to the adsorption of the C$_{60}$ on the Co-substrate. This indicates the hardening the pinning of out-of-plane magnetization due to strong adsorption of C$_{60}$ on Co-substrate. In our previous study on Co-C$_{60}$ in-plane system, we have also seen that introducing a C$_{60}$ layer anisotropy of the system has been increased \cite{mallik2019enhanced}.

\section{Conclusions:} 
The magnetization relaxation and domain wall dynamics in perpendicular magnetic anisotropic thin films of Pt/Co/Pt with and without C$_{60}$ have been discussed in this paper. Bubble domains are observed for both the thin films. However, introducing the C$_{60}$ layer in Pt/Co/Pt thin film reduces the size of the domains. The C$_{60}$-substrate interactions not only modifies on-surface isotropic as well as anisotropic magnetic textures compared to the bare substrates but it causes the reduction in domain size. The engineering in domain size is advantageous for memory storage application point of view. Similarly, the speed of magnetic relaxation ($\tau$ value) is faster for Pt/Co/C$_{60}$/Pt sample. This fast switching has a remarkable impact in device application. Further in this Pt/Co/C$_{60}$/Pt multilayer structure it is expected that a finite amount of interfacial Dzyaloshinskii–Moriya interaction (iDMI) may be observed. In future the quantification of iDMI may be performed to explore the possibility of tuning the iDMI to host any chiral magnetic structures. 

\section*{Conflicts of interest}
There are no conflicts to declare.

\section*{Acknowledgments:}
The authors would like to thank Dr. Tanmay Chabri for technical help during measurement. We also thank Dr. Sougata Mallick, Mr. Gajanan Pradhan, Ms. Esita Pandey and Dr. Braj Bhusan Singh for valuable discussions. We acknowledge Department of Atomic Energy (DAE), and Department of Science and Technology - Science and Engineering Research Board, Govt. of India (DST/EMR/2016/007725, CRG/2019/003237) for the financial support. 

\section*{Author contributions:}
SB has conceived the idea and coordinated the project. SB has designed the experiments and MEA has designed the DFT calculations. SM has prepared the samples. PS and SM have performed the Kerr microscopy experiments. AM has done the DFT calculations. All authors have discussed the results and written the manuscript.

\bibliography{References}
\bibliographystyle{rsc.bst}

\end{document}


\maketitle
\onecolumngrid

\LARGE{\textbf{Supplementary Information \\Effect of Fullerene on domain size and relaxation in a perpendicularly magnetized Pt/Co/C$_{60}$/Pt system}} \\

\large{Purbasha Sharangi,\textit{$^{a}$} Aritra Mukhopadhyaya,\textit{$^{b}$} Srijani Mallik\textit{$^{a}$}, Md. Ehesan Ali$^{\ddag}$\textit{$^{b}$} and Subhankar Bedanta$^{\ast}$\textit{$^{a}$}}\\

\large{\textit{$^{a}$~Laboratory for Nanomagnetism and Magnetic Materials (LNMM), School of Physical Sciences, National Institute of Science Education and Research (NISER), HBNI, P.O.- Bhimpur Padanpur, Via-Jatni, 752050, India. E-mail: sbedanta@niser.ac.in}}\\
\large{\textit{$^{b}$~Institute of Nano Science and Technology, Knowledge City, Sector-81, Mohali, Punjab 140306, India. E-mail: ehesan.ali@inst.ac.in }}\\

{\large\textbf{Structural information:}}
\\In order to investigate the structural information$,$ we have performed X-ray reflectivity (XRR) measurements by a x-ray diffractometer (model SmartLab) manufactured by Rigaku. We have fitted the data by using GenX software.  Figure S1 (a)$,$ (b) show the XRR data and best fits for samples 1 and 2$,$ respectively. Structural parameters such as thickness and roughness of the layers are shown in Table S1.
\\ \large\textbf{Anisotropy calculation:}  
\\ To calculate the magnetic anisotropy, M-H loops have been measured along hard axis (i.e., in-plane) and easy axis (out-of-plane) via SQUID magnetometry on both the samples S1 and S2. The hysteresis loops for samples S1 and S2 are shown in Figure S2 (a) and (b), respectively. We have extracted the effective anisotropy constant ($K_{eff}$) using equation 1. The value of $K_{eff}$ for samples S1 and S2 are $4.3\times 10^{6}$ and $5.1\times 10^{6}$ $J/m^{3}$, respectively. The reduction in coercivity may be due to the anti-parallel alignment between the Co and the magnetic C$_{60}$ moments at the interface. This is something to be disentangled in a future work.

 \begin{equation}
 	K_{eff}= \frac{\mu_{0}H_{S}M_{S}}{2}
 \end{equation}
 
\begin{figure*}
	\centering
	\includegraphics[width=1.0\linewidth]{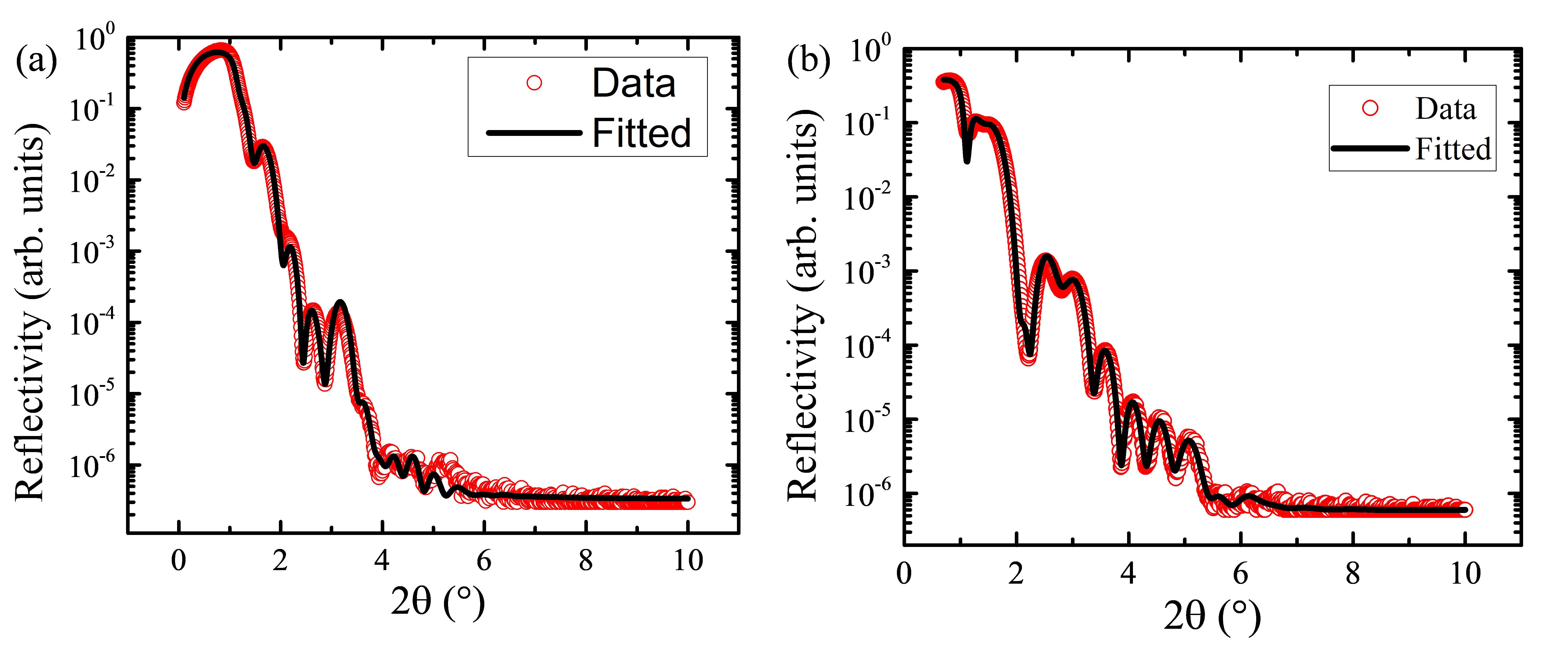}
	\caption{XRR data and the best fits for samples 1 and 2 are shown in (a) and (b)$,$ respectively. The open circles are experimental data and the solid lines represent the best fit using GenX software. The parameters extracted from the best fits are shown in Table S1.}
	\label{figS1}
\end{figure*}
\begin{figure*}
	\centering
	\includegraphics[width=1.0\linewidth]{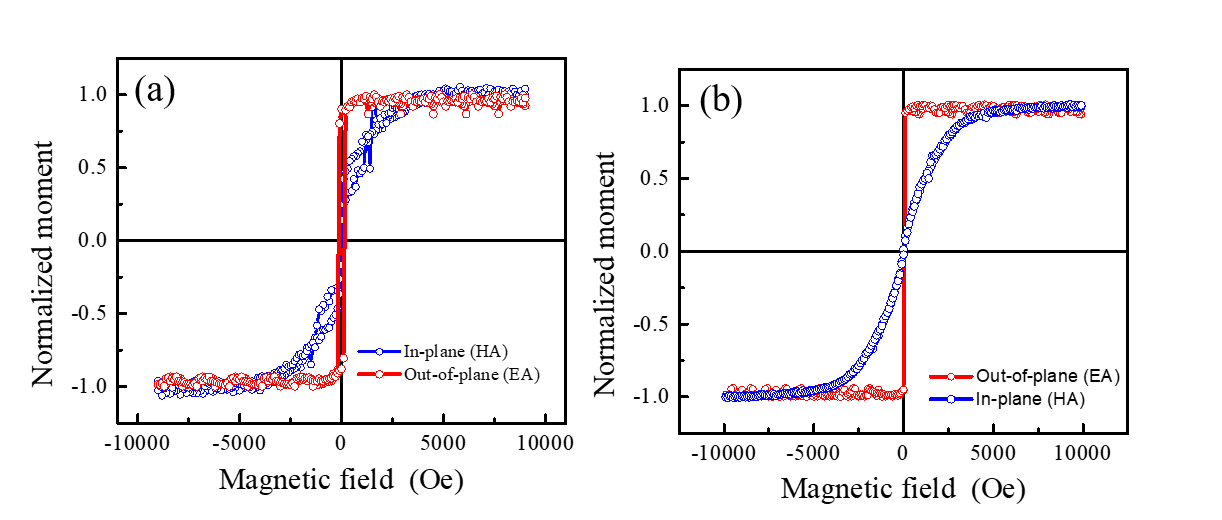}
	\caption{Figure R5: M-H loops along hard axis (in-plane) and easy axis (out-of-plane) via SQUID magnetometry on samples S1(a) and S2 (b).}
	\label{figS2}
\end{figure*}
\begin{table*}
		
	\caption {Parameters obtained from XRR fits }
	\begin{tabular}{|c|c|c|c|c|}
		\hline
		& \multicolumn{2}{c|}{Sample 1}   & \multicolumn{2}{c|}{Sample 2}   \\ \hline
		Layers & Thickness (nm) & Roughness (nm) & Thickness (nm) & Roughness (nm) \\ \hline
		Ta     & 5.8            & 0.75           & 5.5            & 0.56           \\ \hline
		Pt     & 4.5            & 0.74           & 4.1            & 0.57           \\ \hline
		Co     & 0.8            & 0.50           & 0.81           & 0.49           \\ \hline
		C$_{60}$    & -              & -              & 1.7            & 0.77           \\ \hline
		Pt     & 4.8            & 0.698          & 4.6            & 0.69           \\ \hline
	\end{tabular}
\end{table*}

\begin{figure*}
	\centering
	\includegraphics[width=1.0\linewidth]{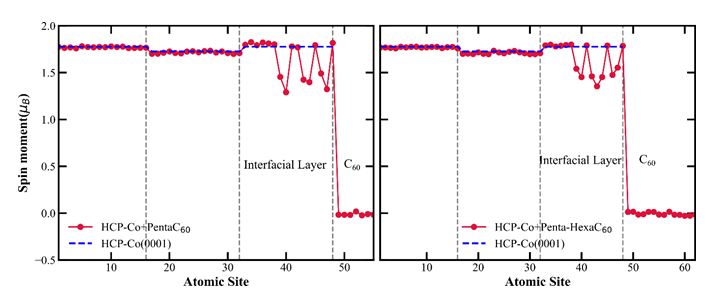}
	\caption{Distribution of spin moments on different atomic sites for HCP-Co (0001) substrate. }
	\label{figS3}
\end{figure*}

\begin{figure*}
	\centering
	\includegraphics[width=1.0\linewidth]{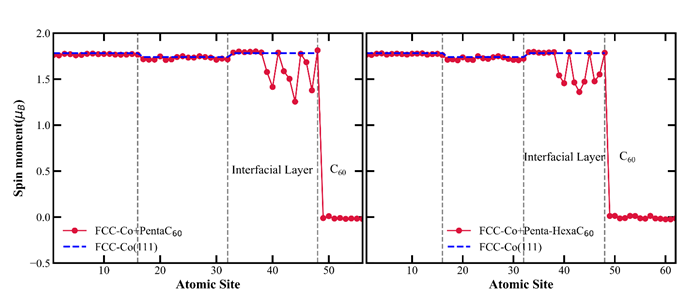}
	\caption{Distribution of spin moments on different atomic sites for FCC-Co (111) substrate.}
	\label{figS4}
\end{figure*}

{\large\textbf{Distribution of spin moment:}}

Figures S3 and S4 depict how the spin moments are distributed over different atomic sites at different Co-C$_{60}$ interface. The top layer of the Co slabs is affected mostly by the spinterface formation. There are mainly three types of spin moments at the considered interfaces. These moments have been used to calculate the exchange interactions.
